# *Vertical Industries Requirements Analysis & Targeted KPIs for Advanced 5G Trials*

Tilemachos Doukoglou, Velissarios Gezerlis
HELLENIC TELECOMMUNICATIONS
ORGANIZATION SA (OTE)
Athens, Greece
{tdouk, vgezerlis}@ote.gr

Konstantinos Trichias
WINGS ICT SOLUTIONS PC.
Athens, Greece
ktrichias@wings-ict-solutions.eu

Nikos Kostopoulos
ERICSSON HELLAS SA
Athens, Greece
nikos.kostopoulos@ericsson.com

Nikos Vrakas, Marios Bougioukos
NOKIA SOLUTIONS AND NETWORKS HELLAS SA
Athens, Greece
{nikos.vrakas, marios.bougioukos}@nokia.com

Rodolphe Legouable
ORANGE SA
Cesson-Sévigné, France
rodolphe.legouable@orange.com

*Abstract* **– Just before the commercial roll-out of European 5G networks, 5G trials in realistic environments have been recently initiated all around Europe, as part of the Phase 3 projects of 5GPPP H2020 program [1]. The goal is to showcase 5G's capabilities and to convince stakeholders about its value-adding business potential. The approach is to offer advanced 5G connectivity to real vertical industries and showcase how it enables them to overcome existing 4G network limitation and other long-standing issues. The 5G EVE H2020 5GPPP project [2] offers cutting-edge 5G end-to-end facilities (in 4 countries) to diversified vertical industry experimenters. The objective is to understand the needs of prominent industries across Europe and to offer tailor-made 5G experience to each and every one of them. This paper contributes to the understanding of vertical services' needs, by offering a thorough and concise vertical requirements analysis methodology, including an examination of the 4G limitations. It also provides real-life values for the targeted KPIs of 3 vertical sectors namely Smart Industry (4.0), Smart Cities / Health and Smart Energy, while assisting market roll-out by prioritizing their connectivity needs.**

## I. INTRODUCTION

The Phase 3 5GPPP H2020 advanced 5G trials with the participation of vertical stakeholders, executed in realistic environments, are expected to verify the superiority of 5G technology compared to its predecessors. The trials pave the way for commercialization of services not possible using existing technologies, together with their concrete business cases. In this context the 5G EVE project [2] aims to be instrumental towards the pervasive roll-out of 5G end-to-end (E2E) networks in Europe. In order to be able to offer the various diversified vertical industries their desired Quality of Service / Experience (QoS / QoE) over the same 5G network, it is paramount to understand in depth their needs and connectivity expectations and to translate them into suitable network configurations and technological features selection. This understanding will lead to the provisioning of an optimum, tailor-made service (5G slice selection of suitable SW and HW components in the Core, Transport and Radio network) per vertical industry, with guaranteed performance metrics. The importance of properly understanding the verticals' needs, and the analysis of their requirements leading to the provisioning of the optimum network slice, has recently been highlighted by the Global System for Mobile Communications Association (GSMA) [3]. The combination of a well-defined use case, with the appropriate evaluation metrics [4], allows for the quantification of the effect of certain technical solutions on the users' QoE as well as the 5G system's performance under realistic, measurable, and verifiable conditions.

Recent European projects have significantly contributed to the requirements' analysis of various vertical industries, thus enabling the definition of targeted Key Performance Indicators (KPIs) per use case. The 5G-TRANSFORMER project has performed a requirements analysis of certain vertical sectors aiming at highly customizable slices through federated virtualized infrastructure [5], while the 5G-PICTURE project delivered a detailed requirements analysis of a disjoint set of verticals focusing on the applicability of the Dis-Aggregated RAN (DA-RAN) concept [6]. Finally, the work carried out in the 5GINFIRE project provided insights regarding the requirements for the coexistence of multiple vertical industries on the same Virtualized Network Functions (VNF) enabled 5G network [7].

The work carried out in the 5G EVE project [8] and presented in this paper aims to extend the state of the art by i) providing a thorough and concise, globally applicable, vertical requirement analysis methodology, ii) benchmarking the performances of 4G and 5G and highlighting the inadequacies of 4G to deal with certain user requirements, iii) providing actual realistic values for the targeted KPIs of three prominent vertical industries (Smart- Industry, City/Health and Energy) originating from vertical sector leaders and iv) prioritizing the needs of the vertical sectors in terms of expected 5G attributes. It has to be noted that the analysis performed in 5G EVE includes more vertical sectors, which was not possible to present here due to space limitation. The full analysis including all six vertical sectors can be found in [8].

## II. NEED FOR REQUIREMENT ANALYSIS IN 5G TRIALS

Requirement Analysis is an essential process to follow when a technology, system or software project emerges or evolves and aims to determine the user expectations upon usage. Optimally, requirements of the analysis should be documented, actionable, measurable, testable, traceable, related to identified business and customer needs or opportunities, and defined to a level of detail sufficient for system design and deployment.

For the mobile telecommunication field, by moving from 4G to 5G technology, new services will be offered that need enhanced capabilities. The services & end-users' growth, leads to network architecture redesign (core, access and radio), taking into account new parameters such as the need for global coverage combined with low latency, high reliability and increased security. Requirement analysis encompasses those tasks that go into determining the needs or conditions to be met for a new or improved service or product offer, while at the same time considering any conflicting requirements by the various stakeholders.

### A. Expectations from Requirement Analysis

With defined requirements we expect a more user-centric design which begins with the understanding of the end-user needs and requirements. The benefits can include increased productivity, enhanced quality of work, reductions in support and training costs, improved user satisfaction, and reduced OPEX. With 5G technologies new service-groups are added such as Internet-of-Things (IoT), virtual and augmented reality (VR and AR), mission critical applications, massive machine-type communication, Gigabit mobile (on the go) connectivity and others which promise to positively impact day-to-day life. Requirements Analysis should lead to service offerings closer to end-user expectation (with improved QoE) while at the same time utilizing existing (4G) and upcoming (5G) network capabilities in a way which reduces unnecessary investment, decreases OPEX and CAPEX and increases resource utilization, while maximizing performance.

### B. Methods for User Requirement Analysis

For an effective user requirement gathering in the context of the 5G EVE project, the following methodology is used: (a) General capabilities of 4G and 5G technology have to be considered, and the differences between them illustrated (b) General 4G/5G requirements for each use case have to be gathered and recorded into tables (c) Special 4G/5G requirements for each use case have to be gathered and recorded also into tables (d) All gathered requirements for each use case have to be illustrated using graphs in correlation with the general 4G/5G capabilities (e) Analysis for each use case requirements can then be extracted from the tables and graphs.

Evaluating the value and impact of vertical implementations is a complex task involving multifaceted analysis. Qualitative methods to explore the effectiveness of vertical implementations include user interviews, observations and open-ended questionnaires targeting all stakeholders in the process. Quantitative measurements of such quantities as data-rate, link capacity, availability, latency and others can be used for evaluation and comparison. Generic and specific user requirements that will result as ad hoc measurements per vertical, shall be evaluated against specific criteria.

## III. THE 5G EVE FACILITY & USE CASE DEFINITION

### A. The 5G EVE End-to-End Facility

The 5G EVE end-to-end facility is comprised of four operational site facilities located in Athens (GR), Turin (IT), Madrid (ES) and France (distributed) and each site facility, is operated by a Telecom operator, i.e. OTE, TIM, Telefonica and Orange, respectively. Each site facility is capable of meeting the requirements of various vertical use cases by deploying their components accordingly. The 5G EVE project is working towards a vertical-agnostic implementation but for development purposes six specific use cases described in Section III.B and in [8], are initially supported. Initial access to verticals for early experimentation is planned for April 2019. From Q1/2020, the 5G EVE facility will also support validation tests from additional vertical use cases proposed by external research projects, including those to be funded under the Horizon 2020 ICT-19 call. A brief description of the 5G EVE facility is presented in this paper, while more information is available in [9] and [10] regarding the architecture and deployment roadmap.

*The Greek 5G EVE site facility* covers a region of Northern Athens, operated by the Greek National Telecommunication Organization (OTE), supported by Ericsson GR, Nokia GR and WINGS ICT SOLUTIONS. The facility is constantly upgraded through R&D operations and will be upgraded with 5G capabilities based on equipment and technology from the two vendors to initially support three vertical use cases (the focus of this paper). *The Italian 5G EVE site facility* is operated by TIM with the help of Ericsson IT, Nextworks and CNIT while Commune di Torino and Trenitalia participate as the vertical users. The facility will be a coherent synthesis of live and laboratory-based experimental environments for the evaluation of 5G features, where Smart Transport and Smart Cities use cases are going to be implemented, supported by the city of Turin.

*The Spanish 5G EVE site facility* is located at IMDEA Networks premises in Leganés/Madrid and it relies on the 5TONIC Open 5G Lab created in 2015 by Telefónica I+D and IMDEA Networks Institute, supported by Ericsson ES, Nokia ES and UC3M, while Telcaria participates as the vertical user. Media & Entertainment, Industry 4.0 and Smart Tourism concepts will be implemented and showcased in this state-of-the-art facility. *The French 5G EVE site facility* is composed of a cluster of four nodes located in different cities. The entrance point of the French cluster is based at Orange in Châtillon, where the ONAP orchestrator manages the other facilities interconnected via VPN IPsec tunnels. Its main feature is that it rests on two main pillars. The first pillar comprises a pre-commercial Nokia 4G/5G E2E network facility in Paris-Saclay, while the second consists of Open Source building blocks and is distributed across several facilities namely, Châtillon-Paris, Rennes (operated by B◇com), and Sophia Antipolis (operated by Eurecom and Nokia Bell Labs). Industry 4.0 and Smart Energy use cases will be addressed in this facility.

These four facilities will provide a seamless single platform experience for vertical experimenters, as the 5G EVE E2E facility, enabled through the implementation of site interconnections, an interworking layer and common management, orchestration and NFV functionalities (see [9] for the 5G EVE E2E facility architecture). The main objective of this facility is to provide services and capabilities that allow verticals to define their experiments across multiple site facilities. The 5G EVE interworking framework is a key element in this process because: i) it provides a common interface to the verticals for defining their experiments and ii) it allows to interconnect 5G EVE sites at different levels, including orchestration, control and data plane.

### B. Selected Use Cases Description

The final edition of the 5G EVE E2E facility will be a vertical-agnostic platform; for development purposes however, the initial version will support six vertical use cases namely Smart –{Industry (4.0), City/Health, Energy (Utilities), Transport, Tourism and Media & Entertainment} (see [8] for detailed analysis). Due to space limitations, this paper focuses on the requirements analysis of the first three.

#### 1) Smart Industry - AGVs use case

Mobile Cloud Robotics (MCR) in a Smart Wireless Logistic (SWL) facility has been identified as an exciting 5G opportunity **Error! Reference source not found.** that will be exploited by Ericsson. Mobile robots will be used to transport goods between various stations or to and from depots. Deploying mobile robots or Automated Guided Vehicles (AGVs) in logistics improves productivity and supports the implementation of effective lean manufacturing. We are enabling realistic MCR scenarios where traditional robots will be replaced by new ones connected to the cloud. These new robots only include low level controls, sensors and actuators while their intelligence is in the cloud meaning they have access to almost unlimited computing power. Altogether, they are more flexible, more usable and more affordable to own and operate. The connection between MCR robots and the cloud is provided through the mobile network and will benefit from the expected 4G and 5G extremely low latency connections. Each AGV includes mainly its own sensors, whose data are collected and sent to the AGV management system, actuators and low-level control logic. The AGV is connected to the AGVs cloud-based management system controlling the AGV and coordinating the operations. RAN indoor dedicated coverage, overlap and coordination for resiliency can be considered (see Fig.1).

A COMAU AGILE1500 AGV with an autonomous remote control is used for shuttling goods between working areas in the warehouse on request. The AGV moves freely using vision and Light Detection and Ranging (LIDAR) to understand its positioning and avoid unexpected obstacles in real time. It is connected via mobile network to a central control running in a local cloud where all the in-tensive control processing is performed. The AGV management systems include several functions from tasks management to trajectory planning, visual navigation and real-time control for the AGV, while a human operator may interact with the control system and the AGV by using an app on a smartphone, tablet or similar device.

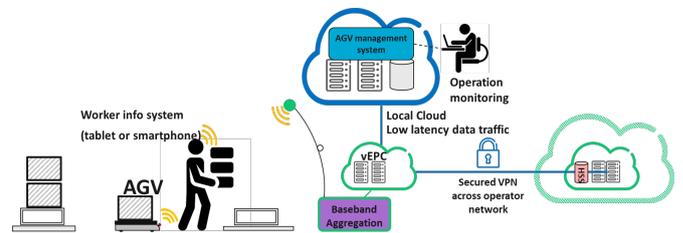

Figure 1: Overall mobile network architecture for an industrial enterprise URLLC application

#### 2) Smart City / Health - Ambulance use case

The 5G "Connected Ambulance" use case advances the emergency ambulance services with their healthcare stakeholders creating improved experiences and outcomes for patients in their care. Communications capabilities that can deliver challenging performance requirements in 5G will be fundamental, as the Connected Ambulance will act as a connection hub (or mobile edge) for the emergency medical equipment and wearables. The convergence of all these technologies will enable the continuous real time streaming patient's vital signs from the moment the emergency ambulance approaches the incident scene, until the arrival to the hospital. Wearables will enable the provision of enhanced patient insights, the real-time video stream to the awaiting emergency department will enable intelligent decision support for the paramedics attending the patient. Life-saving remote assistance on the ambulance can be now feasible when supervised by a specialist connected to the same platform. Such remote assistance requires Ultra High Definition (UHD) video streaming with the minimal latency from the ambulance to the remote site where the specialist is located.

An emergency care platform is provided whereby the patient's vital signs, health stats as well as audio/video data will be wirelessly transmitted real time, whilst on the accident scene and/or on the ambulance, enabling teleconsultation to paramedics, treatment during transport and timely informing of healthcare professionals in next point of definitive care (hospital or other health units). That architecture is comprised briefly of the medical sensors and the UHD enabled camcorder in the ambulance, two gNBs for demonstrating the handover capabilities, the backhaul for reaching the core network, the cloud-based 5G core and the IoT platform.

### 3) Smart Energy / Utilities use case

The integration of an ever-increasing number of distributed electricity sources (renewable energy, farms, house-holds, etc.) into the electricity grid also introduces greater unpredictability of energy production and an increased risk of failures and section isolations (i.e. islanding). Currently, fault detection and management in energy grids, takes place through fibre connectivity among the centralized electricity generation points (e.g. power plants). The move towards Distributed Generators (DG) offers great potential but also makes a fibre-communication monitoring solution prohibitive due to its deployment cost. 5G can enable ultra-fast and ultra-reliable fault detection and management among an extensive number of DGs, with decreased CAPEX and OPEX. Such a fault management system is essential for modern smart grids, enabling immediate reaction to changes in the network thus avoiding unwanted islanding, providing dynamic stability and protection to the network and eventually allowing for the integration of an even greater number of DGs.

One of the largest European energy providers, EDF, participates in 5G EVE and has provided stringent requirements regarding smart grid monitoring operations and ultra-low latency and reliable fault detection and reaction. Through the use of smart metering technology and advanced Machine Learning (ML) techniques, trials over the Greek and French 5G EVE facilities will showcase how 5G connectivity allows for immediate detection and reaction to faults on a distributed smart grid and enables ms-scale smart grid management in an environment of multiple, distributed and heterogeneous energy producers and consumers. The provided 5G requirements and their corresponding analysis are shown in Section IV.D.

## IV. VERTICAL REQUIREMENT ANALYSIS & TARGETED KPIs

Applying the methodology described in Section II on the 5G EVE vertical use cases, enabled the better understanding of the needs of these verticals and provided significant insights into their expectations. The results of the requirements analysis of three vertical use cases is provided in this section.

### A. General capabilities of 4G and 5G

The most common vertical requirements and KPIs that comprised the general 4G and 5G capabilities are shown in Table I below (details in [8]), while a multi-axes radar graph based on these values is depicted in Fig. 2, highlighting the differences in capabilities between 4G and 5G.

TABLE I: 4G/5G CAPABILITIES FOR MAPPING THE VERTICAL'S USE CASES REQUIREMENTS

| | General 4G/5G capabilities | Units | 4G | 5G |
|---|---|---|---|---|
| 1 | E2E Latency (in milliseconds) | msec | 10 | 1 |
| 2 | Speed (in Mbps) - bitrate | Mbps | 400 | 1000 |
| 3 | Reliability (%) | % | 99,9% | 99,999% |
| 4 | Availability (%) | % | 99,9% | 99,999% |
| 5 | Mobility (in m/sec or Km/h) | Km/h | 300 | 500 |
| 6 | Broadband Connectivity (peak demand) | Gbps | 1 | 20 |
| 7 | Network Slicing (Y/N) | Y/N | N | Y |
| 8 | Security (Y/N) | Y/N | Y | Y |
| 9 | Capacity (Mbps/m² or Km²) | Mbps/m² | 0,1 | 10 |
| 10 | Device Density | Dev/Km² | 100K | 1000K |

The selected Use cases can all fit under three main 5G scenarios, namely: i) enhanced Mobile Broadband (eMBB), which needs to support large payloads and high bandwidth, ii) massive Machine Type Communications (mMTC), which needs to support huge number of devices in a base station and iii) Ultra-Reliable Low-Latency Communications (URLLC), which needs to support use cases with a very low latency for services that will require extremely short response times.

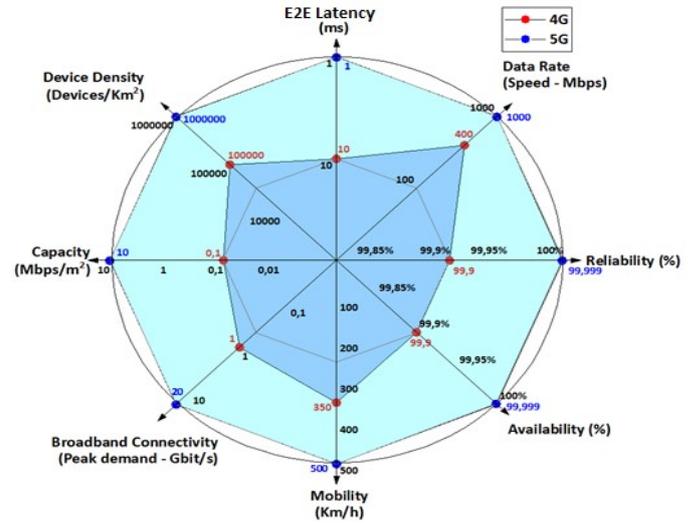

Figure 2: Radar graph for 4G/5G capabilities

### B. Industry 4.0 - Automated Guided Vehicle (AGV) use case.

The 5G EVE AGVs use case belongs to the URLLC 5G scenario. The requirements and targeted KPIs provided by the vertical experimenter are given in TABLE II, while the corresponding multi-axis radar graph is illustrated in Fig. 3.

TABLE II: AGVs USE CASE REQUIREMENTS AND TARGETED KPIs

| | 5G-EVE - Use Cases: direct specific requirements | Units | AGVs use case URLLC | Range Min | Range Max |
|---|---|---|---|---|---|
| | **General Vertical/Use Case Requirement** | | | | |
| 1 | E2E Latency (in milliseconds) - Min/MAX | msec | 10 | 100 | 10 |
| 2 | Speed (in Mbps) - Min/MAX - sustained demand | Mbps | 80 | 80 | 80 |
| 3 | Reliability (%) - Min/MAX | % | 99,999% | 99% | 99,999% |
| 4 | Availability (%) - Min/MAX | % | 99,999% | 99% | 99,999% |
| 5 | Mobility (in m/sec or Km/h) - Min/MAX | Km/hour | 0 | 0 | 0 |
| 6 | Broadband Connectivity (peak demand) | Mbps | 400 | 160 | 400 |
| 7 | Network Slicing (Y/N) | Y/N | Y | N | Y |
| 8 | Security (Y/N) | Y/N | N | N | N |
| 9 | Capacity (Mbps/m^2 or Km^2) | Mbps/m² | 3,2 | 1,28 | 3,2 |
| 10 | Device Density | Dev/Km2 | 40000 | 16000 | 40000 |

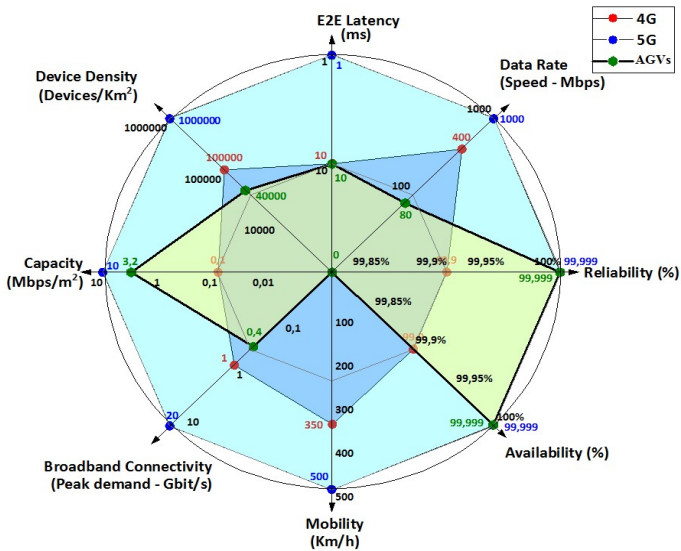

Figure 3: 4G/5G capabilities and AGV Use case - Autonomous guided vehicles in manufacturing environments.

From Table II and the radar graph of AGVs (Fig. 3) it can be easily deduced that the planned service cannot be provided over the existing 4G/LTE networks mainly due to the higher values for the following three requirements: i) capacity, ii) availability and iii) reliability. The AGV use case, is also one of the first to benefit or even be launched as soon as the 5G Networks become commercially available. It appears that this application can even start being offered using existing 4G/LTE network as long as the number of AVGs in a particular space is limited (so as not to exceed the Capacity requirement) and there are lower Reliability and Availability requirements. This might deem the commercial availability of this use case more challenging. Furthermore, if the 5-nines (99.999%) Availability and Reliability requirements cannot be provided in the initial phase of the 5G network, it will potentially delay further the commercial introduction of such a service.

### C. Smart Cities - Smart Ambulance Use case

The 5G EVE Smart Ambulance use case belongs to URLLC, mMTC and eMBB 5G scenarios. The requirements and targeted KPIs provided by the vertical experimenter are shown in TABLE III, while the corresponding multi-axis radar graph is depicted in Fig. 4.

TABLE III: SMART CITIES – SMART AMBULANCE USE CASE REQUIREMENTS AND TARGETED KPIS

| 5G-EVE - Use Cases: direct specific requirements | | Units | Use Case Smart cities Safety and Environment - Smart Ambulance | | | Range | |
|---|---|---|---|---|---|---|---|
| | | | URLLC | mMTC | eMBB | Min | Max |
| General Vertical/Use Case Requirement | | | Smart Cities | | | | |
| 1 | E2E Latency (in miliseconds) - Min/MAX | msec | 5 | <20 ms | <20 ms | 5 | 20 |
| 2 | Speed (in Mbps ) - Min/MAX | Mbps | 0,1 to 1 Mbps | | up to 400 | 25 | 400 |
| 3 | Reliability (%) - Min/MAX | % | 99.99% to 99.9999% | | | 99,99% | 99,9999% |
| 4 | Availability (%) - Min/MAX | % | 99.9 % to 99.99% | | | 99,90% | 99,990% |
| 5 | Mobility (in m/sec or Km/h) - Min/MAX | Km/s | 0km/h-200km/h | 0km/h-200km/h | 0km/h-200km/h | 0 | |
| 6 | Broadband Connectivity (peak demand) | Y/N or Mbps | 25 Mbps | | 200Mbps | 10 | |
| 7 | Network Slicing (Y/N) | Y/N | Y | Y | Y | N | Y |
| 8 | Security (Y/N) | Y/N | Y | Y | Y | N | Y |
| 9 | Capacity (Mbps/m^2 or Km^2) | Mbps/m² | 0,01 to 0,2 Mbps/m² | | | 0,01 | 0,2 |
| 10 | Device Density | Dev/Km2 | | 60K (sensors) | | | 60K |

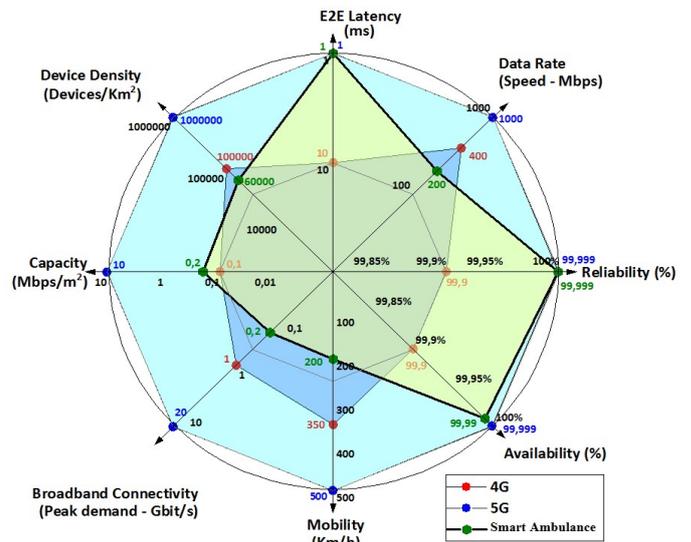

Figure 4: 4G/5G capabilities and Smart Ambulance Use case

It becomes easily apparent that the Smart cities – Smart Ambulance use case requires the enhanced capabilities offered by 5G network in terms of i) Capacity, ii) Latency, iii) Reliability and iv) Availability. The relation of the use case with the Health Industry, more than justifies the required Availability and Reliability. Furthermore, since we have to deal with real time transmission of lossless data streams, low latency is also needed. The materialization of such a service like Smart Ambulance is subject to the capabilities that the 5G networks can offer.

### D. Smart Energy / Utilities use case

The 5G EVE Smart Energy/Utilities use case belongs to URLLC 5G scenarios. The requirements and targeted KPIs provided by the vertical experimenter are illustrated in Table IV, while the corresponding multi-axis radar graph is depicted in Fig. 5.

TABLE IV: SMART ENERGY/UTILITIES USE CASE REQUIREMENTS AND TARGETED KPIS

| 5G-EVE - Use Cases: direct specific requirements | | Units | Smart Energy/Utilities | Priority | Range | |
|---|---|---|---|---|---|---|
| | | | URLLC | | Min | Max |
| General Vertical/Use Case Requirement | | | | | | |
| 1 | E2E Latency (in miliseconds) - Min/MAX | msec | < 5 ms | High | 5 | 30 |
| 2 | Speed (in Mbps ) - Min/MAX | Mbps | <1Mbps | low | | |
| 3 | Reliability (%) - Min/MAX | % | 99,999% | High | 99,999% | |
| 4 | Availability (%) - Min/MAX | % | 99,999% | High | 99,999% | |
| 5 | Mobility (in m/sec or Km/h) - Min/MAX | Km/s | 0 | | | |
| 6 | Broadband Connectivity (peak demand) | Y/N or Mbps | 10 Mbit/s | | | |
| 7 | Network Slicing (Y/N) | Y/N | Y | High | | |
| 8 | Security (Y/N) | Y/N | Y | High | | |
| 9 | Capacity (Mbps/m^2 or Km^2) | Mbps/m² | 0,1 | | | |
| 10 | Device Density | Dev/Km2 | <2000 | medium | | |

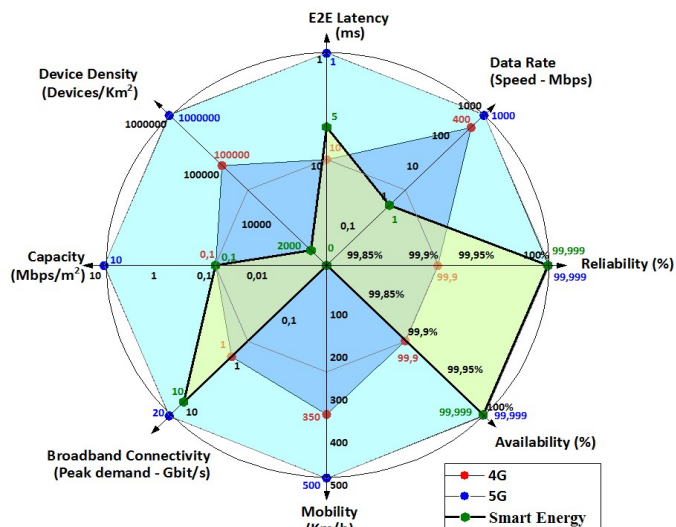

Figure 5: 4G/5G capabilities and Smart Energy/Utilities Use case

From the Smart Energy/Utilities use case requirements (Table IV & Fig.5) it can be seen that only half of the basic/general requirements are being fulfilled by existing 4G/LTE network technologies. Enhanced performance in terms of i) Latency, ii) Broadband Connectivity or Peak Demand, iii) Reliability and iv) Availability are required and can only be provided via the upcoming 5G technology. Energy and Smart Grids are fields in continuous evolution, from monolithic Point to Multi Point unidirectional distribution networks to Multi Point to Multi Point bidirectional (in terms of energy flow) networks (due to introduction of renewable sources and distributed storage). Response time, i.e., to identify faults and rectify and restore energy flow (provisioning) is critical, while the high dependency of other major utility grids on the electricity network more than justifies the high Reliability and Availability requirements. Smart Grid applications are in great demand for the 5G network capabilities.

*E. Extended Analysis*

If we are to prioritize the 5G network capabilities in terms of the vertical stakeholders' demand within the 5G EVE Project context, the capability prioritization presented in the following graph is generated:

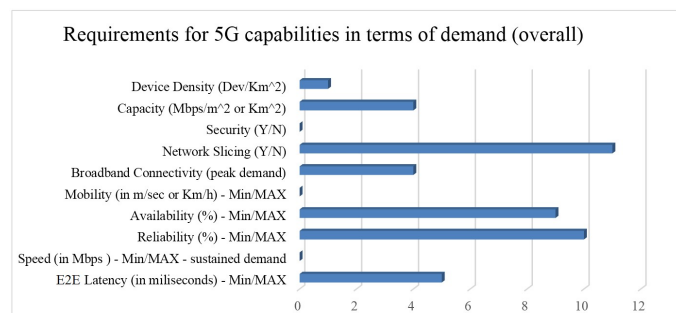

Figure 6: Prioritization of 5G EVE Use cases' 5G Capabilities by the frequency that appears in the requirement tables

The Chart in Fig. 6 indicates that the most desirable capability in 5G networks is Slicing. Although it might not be clear what this capability will offer to the end user, nevertheless it indicates that all, network operators, equipment manufacturers, service providers and end-users require a single network that will fulfil all their needs (no matter how basic and how advanced they might be).

The second most asked-for capability is Reliability. It is an indication that mobile networks should reach a much higher level of reliability (of 5-nines), feature that is now only reserved for fixed-network core-connections and very critical infrastructures (i.e., military etc.).The importance of Reliability and Availability, indicate that service providers and end users alike want the service to be available "always", not only in temporal terms but also in spatial terms. This last requirement (that directly links to network coverage) can potentially make the construction and deployment of the network more expensive. However, cost may potentially be reduced by using network slicing so that the cost is not propagated to the whole infrastructure but just to the pieces involved in those extremely high availability and reliability use cases. In our case, the Availability deals only with the time domain (available 99.999% of the time, which is 5.26 minutes of downtime per year or 25.9 sec per month – on average).

Fourth requirement in priority is Latency, followed by Capacity. It appears that more and more real time applications need to be introduced in the network. Interactive applications also require shorter latency. Finally, the appetite for more capacity is always there since more and more information migrates to online depots/storage and the video resolution (and bandwidth for transmission) requirement also increase continually.

V. CONCLUSIONS

The vertical requirements analysis methodology presented in this paper provides a clear and concise way to perform the requirements analysis and represent the targeted KPIs of vertical industries while also identifying the inadequacies of incumbent technologies, hence highlighting the resolution of problematic areas by 5G technology for each vertical. The disclosed real-life industry KPI values from vertical leaders in smart industry, city / health and energy sectors provide an indication of the expected added-value of 5G by these sectors and assist the telecom operators and vendors in designing their upcoming 5G networks and services, as well as prioritization of their roll-out strategy. As part of the 5G EVE roadmap, vertical industry experimentation will commence on a per site facility basis, targeting to meet the presented KPIs, while during the last year of the project the 5G EVE E2E facility will become available, enabling experimentation across different European facilities and supporting the onboarding of additional vertical industries. The results of these trials will be shared with the scientific and academic communities and benchmarked to the target KPIs as presented here.

ACKNOWLEDGMENT

This work was partly funded by the European Commission under the European Union's Horizon 2020 program – grant agreement number 815074 (5G EVE project). The paper solely reflects the views of the authors. The Commission is not responsible for the contents of this paper or any use made thereof.

## REFERENCES

[1] 5G PPP program, available at https://5g-ppp.eu

[2] 5G EVE project, available at https://www.5G EVE.eu

[3] GSMA 5G Network Slicing Report, "From Verical Industry Requirements to Network Slice Characteristics", Aug. 2018.

[4] Ambler, Scott. "Technical (Non-Functional) Requirements: An Agile Introduction". Agile Modelling. Ambysoft Inc. Retrieved, Oct. 2018.

[5] H2020 5G-TRANSFORMER project, deliverable D1.1 "Report on Vertical Requirements and use cases", Nov. 2017.

[6] H2020 5G-PICTURE project, deliverable D2.1 "5G and Vertical Services, use cases and Requirements", Jan. 2018.

[7] A.Gavras, S.Denazis, H.Hrasnica and C.Tranoris, "Requirements and design of 5G experimental environments for vertical industry innovations", H2020 5GINFIRE project, Global Wireless Summit, Oct. 2017

[8] H2020 5G EVE project, deliverable D1.1 "Requirements Definition & Analysis from Participant Vertical-Industries", Oct. 2018, available at https://www.5g-eve.eu/wp-content/uploads/2018/11/5g-eve-d1.1-requirement-definition-analysis-from-participant-verticals.pdf

[9] H2020 5G EVE project, deliverable D2.1 "Initial detailed architectural and functional site facilities description", Sept. 2018, available at https://www.5g-eve.eu/wp-content/uploads/2018/10/5geve_d2.1-initial-architectural-facilities-description.pdf

[10] H2020 5G EVE deliverable D2.2 "Site facilities planning", Oct. 2018, available at https://www.5g-eve.eu/wp-content/uploads/2018/11/5g-eve-d2.2-site-facilities-planning.pdf

[11] B. Kehoe, S. Patil, P. Abbeel and K. Goldberg, "A Survey of Research on Cloud Robotics and Automation", IEEE transactions on automation science and engineering, 2015.